\documentclass[pra,aps,twocolumn,showpacs,superscriptaddress,nofootinbib]{revtex4}

\newtheorem{theorem}{Theorem}
\newtheorem{lemma}{Lemma}

\begin{document}

\title{Optimal control, geometry, and quantum computing}

\author{Michael A. Nielsen}
\email[http://www.qinfo.org/people/nielsen/blog/]{}
\affiliation{School of Physical Sciences,
The University of Queensland, Queensland 4072, Australia}

\author{Mark R. Dowling}
\affiliation{School of Physical Sciences,
The University of Queensland, Queensland 4072, Australia}

\author{Mile Gu}
\affiliation{School of Physical Sciences,
The University of Queensland, Queensland 4072, Australia}

\author{Andrew C. Doherty}
\affiliation{School of Physical Sciences,
The University of Queensland, Queensland 4072, Australia}

\date{\today}

\begin{abstract}
  We prove upper and lower bounds relating the quantum gate complexity
  of a unitary operation, $U$, to the optimal control cost associated
  to the synthesis of $U$.  These bounds apply for any optimal control
  problem, and can be used to show that the quantum gate complexity is
  essentially equivalent to the optimal control cost for a wide range
  of problems, including time-optimal control and finding minimal
  distances on certain Riemannian, subriemannian, and Finslerian
  manifolds.  These results generalize the results of [Nielsen,
  Dowling, Gu, and Doherty, \emph{Science} \textbf{311}, 1133-1135
  (2006)], which showed that the gate complexity can be related to
  distances on a Riemannian manifold.
\end{abstract}


\maketitle

\section{Introduction} 

Quantum computers have caused great interest due to their potential
use in efficiently solving problems considered intractable on
conventional classical computers~\cite{Shor97a,Nielsen00a}.  Despite
this interest, there is as yet no general framework for constructing
efficient quantum algorithms, nor for proving limitations on the power
of quantum computers.

Recent work~\cite{Nielsen06b,Nielsen06c} has proposed a geometric
approach to quantum computation, based on the observation that finding
quantum circuits of the minimal size required to perform some desired
computation is equivalent to a problem in Riemannian geometry.  More
precisely, the size of the minimum quantum circuit synthesizing a
unitary $U$ is, up to polynomial factors and some technical caveats
(see Section~\ref{sec:examples} for precise statements), equal to the
distance $d(I,U)$ between the identity operation $I$ and $U$,
according to some Riemannian metric.  This equivalence means that
problems in quantum computation can be recast in terms of equivalent
problems in Riemannian geometry.

The results of~\cite{Nielsen06b,Nielsen06c} establish an equivalence
between the number of gates needed to synthesize $U$ and the minimal
distance according to some \emph{specific} Riemannian metric.
However, inspection of the proof in~\cite{Nielsen06b,Nielsen06c} shows
that many of the properties used in the proof are rather generic, and
there are certainly other Riemannian metrics with the same property.
One may therefore ask what is the most general class of Riemannian
metrics that can be connected to gate complexity.  Even more
generally, the problem of finding minimal geodesics in Riemannian
geometry may be viewed as an instance of the problem of optimizing
some cost function in the framework of nonlinear optimal control (see,
e.g.~\cite{Jurdjevic96a}), and it is interesting to ask whether it is
possible to make any general connections between optimal control and
gate complexity.

The purpose of the present paper is to identify a large family of
optimal control problems whose optimal cost is equivalent to the
minimal gate complexity of the desired unitary operation.  As special
cases of our results we obtain the geometric results
of~\cite{Nielsen06b,Nielsen06c}, but also identify many other classes
of optimization problems which can be connected to gate complexity,
including problems from time-optimal control, and from Riemannian,
subriemannian, and Finslerian geometry.  Of course, in some (though
not all) of these examples more straightforward techniques may be used
to relate the optimal cost to quantum gate complexity.  The benefit of
the analysis in the present paper is that it provides a unified and
generalized framework for deriving connections between quantum gate
complexity and optimal control.

By identifying this large family of optimal control problems we
identify the essential features of the geometric problem
in~\cite{Nielsen06b,Nielsen06c} that are responsible for the
equivalence to quantum computation.  We also widen the class of
problems in optimal control which may be analysed in order to obtain
insight into quantum computation.  A considerable body of work has
been done on optimal control in quantum physics (see references later
in the paper), and we hope that the close connection between optimal
quantum control and quantum gate complexity will stimulate further
work on optimal quantum control.

The structure of the paper is as follows.
Section~\ref{sec:background} describes background material on quantum
computing and optimal control theory that is useful later in the
paper.  Section~\ref{sec:bounds} proves a general theorem relating the
optimal cost for a control problem to quantum gate complexity.  In
Section~\ref{sec:examples} we illustrate this theorem through a series
of applications to example problems, including time-optimal control,
and problems from Riemannian, subriemannian and Finsler geometry.
Section~\ref{sec:conclusion} concludes.

\section{Background}
\label{sec:background}

In this section we introduce some background material on quantum
computation (Subsection~\ref{subsec:quantum_computation}) and optimal
control (Subsection~\ref{subsec:optimal_control}) that will be useful
later in the paper.

\subsection{Quantum computation and gate complexity} 
\label{subsec:quantum_computation}

We assume the reader is familiar with basic notions of quantum
circuits (e.g., Chapter~4 of~\cite{Nielsen00a}).  Suppose $U$ is an
$n$-qubit unitary operation.  We define the \emph{exact gate
  complexity} $G(U)$ to be the minimal number of one- and two-qubit
quantum gates required to synthesize $U$ \emph{exactly}, with no
ancilla qubits allowed to assist in the preparation of $U$.  We define
the \emph{approximate gate complexity} $G(U,\epsilon)$ to be the
minimal number of gates required to synthesize some $n$-qubit unitary
operation $V$ satisfying $\|U-V\| < \epsilon$, where $\| \cdot \|$ is
the usual matrix norm.  Once again, no ancilla qubits are allowed to
assist in the synthesis.  Note that in~\cite{Nielsen06b,Nielsen06c}
the notation $m(U)$ was used for the gate complexity.

Our results connect problems in optimal control to the values of
$G(U)$ and $G(U,\epsilon)$.  The typical object of interest in optimal
control is the optimal \emph{cost} $C(U)$ associated to a unitary,
$U$, according to a cost function which is defined precisely below.
Our goal is to identify control problems such that $C(U)$ provides
good lower bounds on the exact gate complexity $G(U)$, and good upper
bounds on the approximate gate complexity $G(U,\epsilon)$.  As a
result, up to polynomial factors the exact synthesis of $U$ without
ancilla must take at least $C(U)$ quantum gates, and $U$ can be
synthesized to accuracy $\epsilon$ using at most $C(U)$ quantum gates.

One might naturally ask if it is possible to extend these results to
prove a similar lower bound involving approximate computation, or an
upper bound involving exact computation.  Parameter counting can be
used to show that a bound of the form $G(U) \leq \mbox{poly}(C(U),n)$
is not possible.  Whether a bound of the form
$\mbox{poly}(C(U),n,1/\epsilon) \leq G(U,\epsilon)$ is possible
remains an open problem.  Fortunately, lower bounds for exact
computation and upper bounds for approximate computation remain of
great interest.

\subsection{Optimal control on $SU(2^n)$} 
\label{subsec:optimal_control}

We now sketch the basic ideas of optimal control theory, following the
standard approach (e.g.,~\cite{Jurdjevic96a}), but omitting
mathematical details regarding smoothness and regularity conditions,
as these are not important for our purposes.

Let $H_1,\ldots,H_m$ be a set of linearly independent matrices in the
Lie algebra $su(2^n)$ of traceless $n$-qubit Hermitian
matrices\footnote{Note that physicists' and mathematicians'
  definitions of Lie algebras differ by a factor of $i$, and so our
  definition of $su(2^n)$ is consistent with the usual mathematical
  definition in terms of traceless \emph{skew}-Hermitian matrics.}.
Our control system is based on Schr\"odinger's equation:
\begin{eqnarray} \label{eq:control_system}
  \frac{dU}{dt} = -i H(t) U(t); \,\,\,\,
  H(t)  \equiv \sum_{j=1}^m h_j(t) H_j,
\end{eqnarray}
where $h(t) = (h_1(t),\ldots,h_m(t))$ is known as the \emph{control
  function}, and we impose the initial condition $U(0) = I$.  Defining
the notation $H_h \equiv \sum_{j=1}^m h_j H_j$, we see that $H(t) =
H_{h(t)}$.  We refer to $H(t)$ as the \emph{control Hamiltonian}
corresponding to the control function $h(t)$.  Note that to any
control Hamiltonian $H(t)$ defined on an interval $[0,T]$ there exists
a unique solution $U(t)$ to Equation~(\ref{eq:control_system}) defined
on the same interval.

In general, the control function $h(t)$ is not allowed to take
arbitrary values, but is constrained to lie in an \emph{allowed
  control region} $A \subseteq R^m$.  We denote the corresponding set
of allowed control Hamiltonians by ${\cal H}_A$.

To complete the specification of the control problem we must also
specify a \emph{cost function}, which is a real-valued function $c : A
\rightarrow R$ on the allowed control region.  Equivalently, it may be
regarded as a function $c : {\cal H}_A \rightarrow R$ on allowed
control Hamiltonians, and it is this viewpoint we shall take most
often.  The cost function allows us to assign a cost to a control
Hamiltonian $H(t)$ defined on an interval $[0,T]$ by $C(H(t)) \equiv
\int_0^T dt \, c(H(t))$.  This allows us to define the \emph{cost of a
  unitary $U$} by $C(U) \equiv \inf_{T,H(t)} C(H(t))$, where we take
the infimum over all intervals $[0,T]$, and over all control functions
$H(t)$ such that $H(t) \in {\cal H}_A$ for all times $t$, and $U(T) =
U$.  Note that in general there is no reason why this infimum should
exist, as there may be no allowed control Hamiltonian $H(t)$ which can
be used to synthesize the desired unitary $U$.  However, if we assume
that the Lie algebra generated by $H_1,\ldots,H_m$ is the full Lie
algebra $su(2^n)$, and that the allowed control region ${\cal H}_A$ is
not trivial, we can ensure that such a control function exists, and so
the infimum is defined~\cite{Jurdjevic72a}.  This condition is known
as the condition that the control system be \emph{bracket-generating}.
Provided reasonable continuity assumptions are made about the cost
function $c(\cdot)$ it can also be shown that the infimum is achieved
for some control function $H(t)$.

The allowed control region ${\cal H}_A$ and the cost function
$c(\cdot)$ jointly specify the control problems we shall be interested
in.  Such control problems are known as \emph{right-invariant control
  problems} on the Lie group $SU(2^n)$, and we shall denote them using
the notation $({\cal H}_A, c)$.

\section{Bounds relating optimal control and quantum gate
  complexity}
\label{sec:bounds}

In this section we develop some general relationships between the cost
function $C(U)$ of a right-invariant control system $({\cal H}_A,c)$
on $SU(2^n)$ and the exact and approximate gate complexities, $G(U)$
and $G(U,\epsilon)$.  Our results generalize and extend the ideas
in~\cite{Nielsen06b,Nielsen06c}.  

\textbf{Splittings:} The key tool we use to relate the cost $C(U)$ to
the gate complexities $G(U)$ and $G(U,\epsilon)$ is an object we refer
to as a \emph{splitting}.  We define splittings in two steps.  First,
we identify a special set ${\cal H}_P \subseteq {\cal H}_A$ of
\emph{preferred} Hamiltonians, which we shall assume are bracket
generating.  Second, we identify a \emph{projection map} $P : {\cal
  H}_A \rightarrow {\cal H}_P$ which takes any allowed Hamiltonian $H$
and projects it onto a preferred Hamiltonian $H_P \equiv P(H)$.  Note
that this can be an arbitrary function, and need not be a projection
in the linear algebraic sense.  We call the pair $({\cal H}_P,P)$ a
\emph{splitting} for the control problem $({\cal H}_A,c)$.

The bounds relating the control cost $C(U)$ to gate complexity will
depend on the particular splitting we choose.  For examples of
``good'' choices of splitting (i.e., choices resulting in fairly tight
bounds between control cost and gate complexity) see the later
examples.  For now we suppose that the choice of splitting has been
fixed, and will show how it can be used to relate the control cost to
gate complexity.

Our construction is rather abstract, and many readers may prefer to
first read the statement of Theorem~\ref{thm:general_relation}, and
then to read section~\ref{sec:examples}, where that theorem is applied
to several example control problems.

\textbf{Relationship between $C(U)$ and $G(U)$:} To express this
relationship we need to define two quantities associated to the
splitting.  The first quantity is the maximal cost of applying any
preferred Hamiltonian, $c_P \equiv \sup_{H \in {\cal H}_P} c(H)$.
Note that we use the subscript $P$ as a mnemonic to indicate that
$c_P$ is a cost associated to the set of preferred Hamiltonians.  The
second quantity is the maximal time $T_P$ required to exactly generate
an arbitrary one- or two-qubit unitary operation by applying
time-dependent preferred Hamiltonians.  

Observe that we can synthesize any one- or two-qubit quantum gate for
a cost at most $c_P T_P$.  Since $U$ can be synthesized exactly using
$G(U)$ one- and two-qubit gates, we deduce the desired bound relating
$C(U)$ and $G(U)$:
\begin{eqnarray}
  C(U) \leq c_P T_P G(U).
\end{eqnarray}
Note that the value of $C(U)$ depends only on the control system,
$({\cal H}_A,c)$, not on the choice of splitting, $({\cal H}_P,P)$.
Thus, different choices of splitting can give rise to different
bounds, and it is necessary to choose the splitting in an intelligent
way to get the best possible bound.  In particular, one should choose
the splitting to minimize the product $c_P T_P$.

\textbf{Relationship between $C(U)$ and $G(U,\epsilon)$:} This
relationship is rather more complex than that between $C(U)$ and
$G(U)$, and is expressed in terms of four quantities associated to the
splitting.  The first quantity is a ratio defined by\footnote{Note
  that here and elsewhere we write $\max$ and $\min$ rather than
  $\sup$ and $\inf$.  Our proofs are easily modified for the case when
  (for example) the maximum is not defined, but this does make the
  discussion less transparent, and so we have avoided it.} $R \equiv
\max_{H \in {\cal H}_A} \|H-H_P\|/c(H)$.  The second quantity is the
maximum matrix norm $N_P \equiv \max_{H \in {\cal H}_P} \|H \|$ of any
preferred Hamiltonian.

The third quantity requires a more complex explanation.  Suppose
$\Delta > 0$ and $\delta > 0$.  We define a $\Delta$-averaged
Hamiltonian to be a Hamiltonian $\overline H$ which can be written in
the form $\overline H = \int_0^\Delta dt \, H(t)$ for some Hamiltonian
control function which remains in the preferred set, $H(t) \in {\cal
  H}_P$.  We define the $\Delta$-averaged unitaries to be the set of
unitary operations which can be written in the form $\exp(-i \overline
H)$ for some $\Delta$-averaged Hamiltonian $\overline H$.  We define
$g(\Delta,\delta)$ to be the maximum number of one- and two-qubit
gates required to approximate an arbitrary $\Delta$-averaged unitary
to an accuracy better than $\delta$ in matrix norm.

The fourth quantity is the minimal cost associated to any allowed
Hamiltonian, $c_A \equiv \min_{H \in {\cal H}_A} c(H)$.  This quantity
arises in our proof as a way of getting a bound on the time $T$
associated to the optimal Hamiltonian control $H(t)$.  The argument is
to observe that $C(U) = \int_0^T dt \, c(H(t))\geq T c_A$, and so $T
\leq C(U)/c_A$.

With these quantities defined, we can relate $C(U)$ and
$G(U,\epsilon)$.  The first step is to take the Hamiltonian control
$H(t)$ which achieves the optimal control cost $C(U)$, and to form the
corresponding projected Hamiltonian $H_P(t) \equiv P(H(t))$.  We
suppose $H_P(t)$ generates a unitary $U_P$, and aim to show that $U_P$
is a pretty good approximation to $U$.  As in the proof of Lemma~1 in
the supporting online materials for~\cite{Nielsen06c}, we can apply
the triangle inequality repeatedly to obtain:
\begin{eqnarray}
  \|U - U_P \| & \leq & \int_0^T dt \| H(t)-H_P(t) \|.
\end{eqnarray}
The definition of the ratio $R$ ensures that $\|H-H_P\| \leq R c(H)$
for all $H$, and thus:
\begin{eqnarray}
 \int_0^T dt \| H(t)-H_P(t) \| \leq R \int_0^T dt \, c(H) = R C(U).
\end{eqnarray}
Putting these inequalities together we obtain $\|U-U_P \| \leq R
C(U)$.  Intuitively, provided the control problem and splitting are
such that $R$ is much smaller than $1/C(U)$, we ensure that $U$ and
$U_P$ will be quite close.

In the next step of the proof we discretize the evolution according to
$H_P(t)$, and show that it can be approximated by a suitable sequence
of $\Delta$-averaged Hamiltonians.  The key to doing this is the
following lemma, which appeared as Lemma~2 in~\cite{Nielsen06c}.  We
have made some minor notational changes to the statement of the lemma,
but the essential content of the lemma, and the proof, which is an
easy application of the Dyson operator expansion, is unchanged.

\begin{lemma} Let $V$ be an $n$-qubit unitary generated by applying
  a time-dependent Hamiltonian $H_P(t) \in {\cal H}_P$ over a time
  interval $[s,s+\Delta]$.  Then defining the corresponding
  $\Delta$-averaged Hamiltonian $\overline H \equiv \int_s^{s+\Delta}
  dt \, H(t)$ we have:
\begin{eqnarray}
  \| V - e^{-i\overline{H} \Delta} \| \leq 2 (e^{N_P \Delta}-1-N_P \Delta) =
  O(N_P^2 \Delta^2), \nonumber \\
\end{eqnarray}
where $N_P$ is the maximum matrix norm of any preferred Hamiltonian,
as defined earlier.
\end{lemma}

To apply this lemma, we divide the time interval $[0,T]$ up into a
large number $N$ of time intervals each of length $\Delta = T/ N$. Let
$U_P^j$ be the unitary operation generated by $H_P(t)$ over the $j$th
time interval.  Let $U_M^j$ (the unitary corresponding to the mean
Hamiltonian) be the unitary operation generated by the
$\Delta$-averaged Hamiltonian over the corresponding time interval.
Then the lemma implies that $\| U_P^j - U_M^j \| \leq O(N_P^2
\Delta^2)$.  By assumption, we can then synthesize a unitary operation
$U_A^j$ using at most $g(\Delta,\delta)$ one- and two-qubit gates, and
satisfying $\| U_M^j - U_A^j \| \leq \delta$.  We define $U_A$ (the
actual unitary to be synthesized by our gate sequence) to be the
result of applying the unitaries $U_A^j$ in sequence.  Note that $U_A$
can be generated using $N g(\Delta,\delta) = T
g(\Delta,\delta)/\Delta$ one- and two-qubit quantum gates.

Repeated application of the triangle inequality, substitution of the
inequalities obtained above, and using the fact that $N = T/\Delta$,
yields:
\begin{eqnarray} 
 & & \| U-U_A \| \\
 & \leq & \|U-U_P\| + \| U_P - U_A \|
\\
 & \leq & R C(U) + \sum_{j=1}^N \| U_P^j - U_A^j \| \\
 & \leq & R C(U)
 + \sum_{j=1}^N \left( \|U_P^j-U_M^j \| + \|U_M^j-U_A^j\| \right) \\
 & \leq & R C(U) + O(N_P^2 T \Delta)
 + \frac{T}{\Delta} \delta.
\end{eqnarray}
Substituting the bound on $T$ obtained earlier, $T \leq C(U)/c_A$, we
deduce that we can synthesize an operation $U_A$ satisfying
\begin{eqnarray}
  \|U-U_A\| \leq R C(U) + O\left(\frac{N_P^2 C(U) \Delta}{c_A}\right) +
  \frac{C(U)\delta}{c_A \Delta} \nonumber \\
\end{eqnarray}
using $C(U) g(\Delta,\delta)/ c_A \Delta$ one- and two-qubit gates.

Summing up, we have the following theorem:

\begin{theorem} {} \label{thm:general_relation} 
  Consider a control problem $({\cal H}_A,c)$ and a splitting $({\cal
    H}_P, P)$ for that problem. Then we have:
  
  (1) Let $c_P \equiv \max_{H \in {\cal H}_P} c(H)$ be the maximal
  cost of any preferred Hamiltonian, and suppose $T_P$ is the maximal
  time required to generate an arbitrary one- or two-qubit unitary
  operation using preferred Hamiltonians.  Then:
  \begin{eqnarray}
    C(U) \leq c_P T_P G(U).
  \end{eqnarray}
  
  (2) Let $R \equiv \max_{H \in {\cal H}_A} \|H-H_P\|/c(H)$, $N_P
  \equiv \max_{H \in {\cal H}_P} \|H \|$, $c_A \equiv \min_{H \in
    {\cal H}_A} c(H)$.  Suppose that if $\overline H$ is a
  $\Delta$-average of Hamiltonians in ${\cal H}_P$, i.e., can be
  written in the form $\overline H = \int_0^\Delta dt \, H(t)$ for
  some Hamiltonian control function $H(t)$ which remains in the
  preferred set, then the corresponding unitary $\exp(-i \overline H)$
  can be simulated to an accuracy $\delta$ using a number of gates
  $g(\Delta,\delta)$.  Then we can synthesize an operation $U_A$
  satisfying
  \begin{eqnarray}
  \|U-U_A\| \leq R C(U) + O\left(\frac{N_P^2 C(U) \Delta}{c_A}\right) +
  \frac{C(U)\delta}{c_A \Delta} \nonumber \\
\end{eqnarray}
using $C(U) g(\Delta,\delta)/ c_A \Delta$ one- and two-qubit gates.

\end{theorem}

We stress that this theorem does not necessarily give tight
connections between optimal costs and gate complexity.  Finding such
connections depends on making an appropriate choice of the cost
function, and of the splitting.  However, the examples in the next
section will show that such choices can be made for a wide variety of
interesting cost functions.

\section{Examples} 
\label{sec:examples}

We will now describe a sequence of examples illustrating
Theorem~\ref{thm:general_relation}.  These examples are not
exhaustive, but illustrate the wide range of situations in which
Theorem~\ref{thm:general_relation} can be used to relate problems of
optimal control and quantum gate complexity.  

Note that in each of the examples described in the present section, we
are imagining that there is a \emph{family} $U = U_n$ of unitary
operations, one for each value of $n$, acting on $n$ qubits.
Correspondingly, in each of our examples we will describe an entire
family of cost functions and splittings, one for each value of $n$.
Our goal is to prove results of the form $\mbox{poly}(C(U),n) \leq
G(U)$ and $G(U,\epsilon) \leq \mbox{poly}(C(U),n,1/\epsilon)$ for
suitable polynomial functions.

\textbf{Subriemannian metric:} Suppose the allowed Hamiltonians ${\cal
  H}_A$ are of the form $H = \sum_\sigma h_\sigma \sigma$, where the
sum is restricted to be over Pauli sigma matrices containing only one-
and two-qubit terms, and we require that $\sum_\sigma h_\sigma^2 = 1$.
We define the cost function by $c(H) \equiv \sqrt{\sum_\sigma
  h_\sigma^2}$ so for allowed Hamiltonians we have $c(H) = 1$.  This
cost function $C(U)$ is an example of the distance associated to a
subriemannian metric~\cite{Montgomery02a}, and the problem of finding
$C(U)$ is that of finding the minimal length geodesics on a
subriemannian manifold.  We choose the splitting to be trivial, with
${\cal H}_P = {\cal H}_A$ and $P(H) = H$.

With this control problem $({\cal H}_A,c)$ and splitting $({\cal
  H}_P,P)$, we may apply part~(1) of Theorem~1.  In that notation, it
follows immediately from the definitions that $c_P = 1$ and $T_P$ is a
constant of order one, independent of the number of qubits, $n$.  Thus
$C(U) \leq T_P G(U)$, and so, up to a constant factor, the
subriemannian distance $C(U)$ provides a lower bound on the exact gate
complexity $G(U)$.

To apply part~(2) of Theorem~\ref{thm:general_relation}, note that we
have $R = 0$ and $c_A = 1$, again directly from the definitions.  It
follows from elementary norm inequalities that $N_P =
O(n)$\footnote{$\|H\| \leq \sum_\sigma |h_\sigma| \leq (\sqrt{3}/2) n
  \sqrt{ \sum_\sigma h_\sigma^2} = (\sqrt{3}/2) n$, $\forall H \in {\cal
    H}_P$. The second inequality follows from $\|\vec{v} \|_1
  \leq \sqrt{d} \|\vec{v} \|_2$ where $d$ is the dimension
  of the real vector $\vec{v}$, $\|\vec{v} \|_1 = \sum_{i=1}^d |v_i|$
  and $\|\vec{v} \|_2 = \sqrt{\sum_{i=1}^d v_i^2}$. In our case 
  $d=9n(n-1)/2+3n$, the number of one- and two-qubit terms.  }. To
understand the
behaviour of $g(\Delta,\delta)$, suppose that $\overline H =
\int_0^\Delta dt \, H(t)$ is a $\Delta$-averaged Hamiltonian over
Hamiltonians in ${\cal H}_P$.  Lemma~3 in~\cite{Nielsen06c} implies
that $\exp(-i \overline H)$ can be simulated to an accuracy of order
$O(n^4 \Delta^3)$ using $O(n^2 / \Delta)$ gates.  Thus $g(\Delta,
O(n^4 \Delta^3)) \leq O(n^2/\Delta)$.  We deduce that we can
synthesize an operation $U_A$ satisfying
\begin{eqnarray}
  \| U - U_A \| \leq O(C(U) n^2 \Delta) + O(C(U) n^4 \Delta^2)
\end{eqnarray}
using $O(C(U) n^2 / \Delta^2)$ gates.  It follows that by choosing
$\Delta$ appropriately, we can synthesize a good approximation to $U$
using a number of gates that scales in a fashion comparable to $C(U)$.
To see this, let $\Delta = \epsilon / n^2 C(U)$.  Then we see that we
can synthesize an operation $U_A$ satisfying $\|U-U_A\| \leq
O(\epsilon)$ using $O(C(U)^3 n^6 / \epsilon^2)$ gates.  It follows
that:
\begin{eqnarray}
  G(U,\epsilon) \leq O(C(U)^3 n^6/\epsilon^2),
\end{eqnarray}
which is the required result --- $G(U,\epsilon)$ scales as no more than a
polynomial in $C(U)$, $n$ and $1/\epsilon$.

\textbf{Time-optimal control:} If $c(H) = 1$, then $C(U)$ is the
minimal time taken to generate $U$ using control Hamiltonians in the
allowed control region, ${\cal H}_A$.  This is known as the
\emph{time-optimal} control problem.  A common variant of the
time-optimal control problem is to constrain the set of allowed
controls so that $h_1(t) = 1$, i.e., so that the Hamiltonian $H_1$ is
always being applied.  This is known as the time-optimal control
problem with \emph{drift}, and $H_1$ is known as the \emph{drift
  Hamiltonian}.  The time-optimal control problem in quantum physics
has received considerable attention; see,
e.g.,~\cite{Khaneja01a,SchulteHerbruggen05a,Boscain05a,Agrachev06a,Carlini06a}
for recent work, and further references.  Of particular interest in
this context is work such as~\cite{SchulteHerbruggen05a}, which
studies the time complexity of various quantum computing primitives,
such as the quantum Fourier transform, and applies powerful tools from
optimal control theory such as the Pontryagin maximum
principle~\cite{Pontryagin62a} (see, e.g.,~\cite{Jurdjevic96a}) to
obtain time-optimal implementations of these primitives.

The time-optimal control problem with drift takes a particularly
simple and appealing form in the case where there are only two terms
in the control Hamiltonian, i.e., $H = H_1 + h(t) H_2$, and it is this
case we shall focus on; analogous results can also be proved for other
time-optimal control problems using essentially the same ideas.  We
will assume that the control region is such that the allowed range of
values for $h(t)$ is $|h(t)| \leq 1$.  \emph{A priori} it is not
obvious that it is possible to find examples of Hamiltonians $H_1$ and
$H_2$ which are bracket-generating.  However, it follows from results
of \cite{Lloyd95a,Weaver00a} (c.f.~\cite{Deutsch95a}) that if we
choose $H_1$ and $H_2$ at random, then with probability one they will
be bracket-generating.  Of course, this does not mean that they are
universal for quantum computation in the usual sense.  It may take
such a $H_1$ and $H_2$ exponential time to generate standard quantum
gates such as the controlled-{\sc not}, or even single-qubit
unitaries.  Conversely, it may not be possible to efficiently simulate
$H_1$ and $H_2$ in the standard quantum gate model of computation.

We will now provide examples of families of Hamiltonians $H_1$ and
$H_2$ such that the time-optimal control cost scales as a polynomial
in the quantum gate complexity.  The key to this is the following
theorem, which is of independent interest:

\begin{theorem} {} \label{thm:paired_universality}
  There is a family of $n$-qubit Hamiltonians $H_1$ and $H_2$ such
  that: (1) any one- or two-qubit unitary gate can be synthesized
  exactly in a time bounded above by a value that scales as a
  polynomial in $n$; and (2) using one- and two-qubit gates we can
  simulate any unitary of the form $\exp(-i \Delta(H_1+\alpha H_2))$
  (with $|\alpha| \leq 1$) to an accuracy $\delta$ using
  $g(\Delta,\delta) = O(p(n) \Delta^2 / \delta)$ one- and two-qubit
  gates, for some polynomial $p(n)$.
\end{theorem}

\textbf{Proof (outline):} We choose $H_1$ to be a Hamiltonian acting
on the first two qubits in a manner specified more precisely below.
We choose $H_2$ so that $\exp(-i H_2)$ permutes qubits $2$ through $n$
by a cyclic displacement, i.e., the state of qubit $2$ becomes the
state of qubit $3$, the state of qubit $3$ becomes the state of qubit
$4$, and so on, with the state of qubit $n$ becoming the state of
qubit $2$.

With these choices, conclusion~(2) follows from standard quantum
simulation techniques for simulating a sum of Hamiltonians, and the
observation that the Hamiltonians $H_1$ and $H_2$ can both be
efficiently simulated (the latter using the quantum Fourier
transform~\cite{Shor97a,Nielsen00a}).  

Conclusion~(1) requires a little more effort.  In particular, note
that using $H_1$ and $H_2$ we can simulate the Hamiltonian $\exp(-i
H_2) H_1 \exp(i H_2) = \tilde H_1$, where the tilde denotes that
$\tilde H_1$ is the same Hamiltonian as $H_1$, but now acts on qubits
$1$ and $3$.  It can now be verified numerically or by hand that for
many choices of two-qubit Hamiltonian $H_1$, the Hamiltonians $H_1$
and $\tilde H_1$ generate the full Lie algebra on qubits one, two, and
three\footnote{Examples of this phenomenon were found numerically by
  the present author and H.~L.~Haselgrove~\cite{Nielsen03d}.  C.~Hill
  and Haselgrove~\cite{Hill05a} have recently constructed rather more
  elegant examples demonstrating essentially the same phenomenon as
  described in this theorem, but making use of a relatively simple
  (and more physically plausible) two-body Hamiltonian in place of
  $H_2$, which involves complex many-body terms.}.  As a result, in
constant time we can generate an arbitrary unitary operation on qubits
one, two and three.  Conjugating repeatedly by $\exp(-i H_2)$ we can
use this to generate an arbitrary unitary on qubits $1$ and $j$, where
$j$ is any qubit.  Standard techniques then suffice to efficiently
generate an arbitrary unitary on any pair of qubits.  \textbf{QED}

Suppose we consider the time-optimal control problem where $H_1$ and
$H_2$ have been chosen as in Theorem~\ref{thm:paired_universality}.
As in the subriemannian case we again choose the trivial splitting,
${\cal H}_P = {\cal H}_A$ and $P(H) = H$.  Applying part~(1) of
Theorem~\ref{thm:general_relation}, we see that $c_P = 1$ and $T_p
\leq q(n)$, for some polynomial $q(n)$.  As a result, we have $C(U)
\leq q(n) G(U)$.

Applying part~(2) of Theorem~\ref{thm:general_relation}, we have $R =
0$, $N_P = O(1)$, $c_A = 1$ and $g(\Delta,\delta) = O(p(n)
\Delta^2/ \delta)$, for some polynomial $p(n)$.  As a result, we
conclude that it is possible to synthesize a unitary $U_A$ satisfying
\begin{eqnarray}
  \|U-U_A\| \leq O(C(U) \Delta) + O\left( \frac{C(U) \delta}{\Delta} \right)
\end{eqnarray}
using $O(C(U) p(n)\Delta / \delta)$ gates.  Setting $\Delta =
\epsilon/ C(U)$ and $\delta = \epsilon^2/ C(U)^2$, we see that we can
synthesize a unitary $U_A$ satisfying $\| U - U_A\| \leq O(\epsilon)$
using $O(C(U)^2 p(n)/\epsilon)$ gates, and so we conclude that
\begin{eqnarray}
  G(U,\epsilon) \leq O(C(U)^2p(n)/\epsilon),
\end{eqnarray}
which is the desired polynomial scaling.

\textbf{Riemannian metric:} We now analyze the metric considered
in~\cite{Nielsen06c}, and show how to recover the results
of~\cite{Nielsen06c}.  This is our first example which makes use
of a nontrivial splitting.  Expanding the control Hamiltonian as $H =
\sum_\sigma h_\sigma \sigma$, where the sum is over all $n$-qubit
Pauli matrices, the cost function of~\cite{Nielsen06c} (which is just
the norm associated to the metric) is defined by:
\begin{eqnarray}
  c(H) \equiv \sqrt{\sum_\sigma' h_\sigma^2 + p^2 \sum_{\sigma}'' h_\sigma},
\end{eqnarray}
where the primed sum is over one- and two-qubit Pauli terms, and the
double primed sum is over three- and more-qubit Pauli terms.  The
parameter $p$ is a penalty whose value we set later.  ${\cal H}_A$ is
defined to contain all those Hamiltonians such that $c(H) = 1$.  For
the splitting, we choose the set of preferred Hamiltonians ${\cal
  H}_P$ so that it contains all Hamiltonians containing just one- and
two-qubit terms.  The projection $P$ takes an arbitrary Hamiltonian,
and eliminates all terms in the Pauli expansion except the one- and
two-qubit terms, i.e., it takes $\sum_\sigma' h_\sigma \sigma +
\sum_\sigma'' h_\sigma \sigma$ to $\sum_\sigma' h_\sigma \sigma$.  In
the language of part~(1) of Theorem~\ref{thm:general_relation} we have
$c_P = 1$ and $T_P$ is a constant, and so $C(U) \leq T_P G(U)$, i.e.,
the control cost is a lower bound on the gate complexity $G(U)$, to
within a constant factor.

Next, we evaluate the quantities defined in part~(2) of
Theorem~\ref{thm:general_relation}.  To evaluate $R$, observe that the
Hamiltonian $H$ achieving the maximum must contain only terms which
are three- or more-body.  Thus:
\begin{eqnarray}
  R =
  \frac{\| \sum_\sigma'' h_\sigma \sigma \|}{p \sum_\sigma'' h_\sigma^2}
  \leq \frac{\sum_\sigma'' |h_\sigma|}{p \sum_\sigma'' h_\sigma^2}
  \leq \frac{2^n}{p},
\end{eqnarray}
where the first inequality follows from the triangle inequality, and
the second inequality follows from the Cauchy-Schwarz inequality.  For
the same reasons as in the subriemannian case, $N_P = O(n)$, $c_A = 1$
and $g(\Delta, O(n^4 \Delta^3)) \leq O(n^2/\Delta)$.

Applying part~(2) of Theorem~\ref{thm:general_relation} we deduce that
we can synthesize an operation $U_A$ satisfying
\begin{eqnarray}
  \| U - U_A \| \leq \frac{2^n}{p} C(U) + 
  O(C(U) n^2 \Delta) + O(C(U) n^4 \Delta^2) \nonumber \\
\end{eqnarray}
using $O(C(U) n^2 / \Delta^2)$ gates.  Again we choose $\Delta =
\epsilon / n^2 C(U)$.  Then we see that we can synthesize an operation
$U_A$ satisfying $\|U-U_A\| \leq 2^n C(U)/ p + O(\epsilon)$ using
$O(C(U)^3 n^6 / \epsilon^2)$ gates.  Standard results on
universality (see, e.g.,~\cite{Shende04a} and references therein)
imply that $C(U) \leq O(4^n)$ for all unitaries $U$, and so by
choosing $p = 8^n/\epsilon$ we obtain
\begin{eqnarray}
  G(U,\epsilon) \leq O(C(U)^3 n^6/\epsilon^2),
\end{eqnarray}
which is the desired polynomial scaling.

\textbf{Other control problems:} It is not difficult to generate many
other examples of optimal control problems whose cost scales in
essentially the same way as the gate complexity.  An example is the
following cost function that was conjectured in~\cite{Nielsen06b} to
be equivalent to the gate complexity:
\begin{eqnarray}
  c\left( \sum_\sigma h_\sigma \sigma \right) \equiv \sum_\sigma' |h_\sigma|
  + p \sum_\sigma'' |h_\sigma|.
\end{eqnarray}
Once again, $p$ is a penalty parameter that we shall choose to be
large.  We define ${\cal H}_A$ to consist of all Hamiltonians such
that $c(H) = 1$.  We define the splitting as for the Riemannian metric
considered above, setting ${\cal H}_P$ to be those Hamiltonians in
${\cal H}_A$ containing only one- and two-qubit terms, and the
projection $P$ to remove all three- and more-qubit terms from the
Pauli expansion.  A similar analysis to the Riemannian case allows us
to relate the cost $C(U)$ to the gate complexity.  The only
significant difference is in the evaluation of $R$, where we obtain $R
\leq 1/p$, and thus it is possible in this case to choose more modest
values of $p$ and still achieve a close relationship between the
scaling of the cost and of the gate complexity.

\section{Conclusion}
\label{sec:conclusion}

We have proved a general theorem relating quantum gate complexity to
the optimal control cost for an arbitrary control problem.
Application of the theorem depends on the use of a tool known as a
\emph{splitting}, which must be chosen appropriately in order to
obtain good bounds.  We have illustrated this theorem with examples
showing that quantum gate complexity is essentially equivalent to the
optimal control cost for problems including time-optimal control and
finding minimal distances on certain Riemannian, subriemannian, and
Finslerian manifolds.  It is possible to improve the scaling in many
of these results with a more refined use of the Dyson operator
expansion~\cite{Sakurai94a} and Suzuki-Trotter type
formulas~\cite{Suzuki90a}, and it would be interesting to determine
what the optimal bounds are.  It also seems likely that the results
can be further generalized using tools more sophisticated than the
notion of a splitting that we have introduced.  However, the most
important direction of future work will be to better understand the
optimal cost for specific choices of control problem, and what it
implies for quantum gate complexity.

\acknowledgments
Thanks to Lyle Noakes for his encouragement, and for emphasizing the
importance of isolating the essential features of optimal control
problems responsible for the equivalence to quantum gate complexity.

\bibliography{../../mybib}

\end{document}